\documentclass[conference]{IEEEtran}
\IEEEoverridecommandlockouts
\usepackage{filecontents,lipsum}
\ifCLASSOPTIONcompsoc

\else
\usepackage{cite}
\fi
\ifCLASSINFOpdf
\usepackage[pdftex]{graphicx}
\else
\fi
\usepackage{amsmath}
\usepackage{amssymb}
\usepackage{array}
\usepackage[caption=false,font=footnotesize]{subfig}
\graphicspath{ {Figures/} }
\usepackage{float}
\usepackage{fontawesome}
\usepackage{multirow}
\usepackage{color}
\usepackage{lipsum}
\usepackage{mathtools}
\usepackage{cuted}
\usepackage{soul}
\usepackage{color, xcolor}
\usepackage{lipsum}
\usepackage{mathtools}
\usepackage{nomencl}
\makenomenclature

\renewcommand{\figurename}{Fig.}
\usepackage{stfloats}
\usepackage{calc}
\newlength{\depthofsumsign}
\newlength{\totalheightofsumsign}
\newlength{\heightanddepthofargument}

\usepackage{nicematrix}
\usepackage{arydshln}
\usepackage{subfig}
\usepackage{lipsum}
\usepackage{bbding}

\usepackage{amsmath}

\usepackage{caption}
\usepackage{subcaption}
\usepackage{amsthm}
\usepackage{graphicx}
\usepackage{subfig}

\renewcommand\IEEEkeywordsname{Keywords}
\renewcommand{\figurename}{Fig.}

\usepackage{xfrac}
 \usepackage{algorithm}
 \usepackage{algpseudocode}
 \usepackage{midfloat}
\begin{document}

\title{Physics-informed Convolutional Neural Network for Microgrid Economic Dispatch\\
\thanks{This work was in part under support from the Department of Defense, Office of Naval Research award number N00014-23-1-2602 and was financed (in part) by a grant from the Commonwealth of Pennsylvania,
Department of Community and Economic Development, through the Pennsylvania Infrastructure
Technology Alliance (PITA). Xiaoyu Ge is a PhD student and Javad Khazaei is an assistant professor in the electrical and computer engineering (ECE) department at Lehigh University, Bethlehem, PA. (emails: xig620@lehigh.edu , \& khazaei@Lehigh.edu).}
}

\author{
\IEEEauthorblockN{Xiaoyu Ge,~\IEEEmembership{Student Member,~IEEE} and Javad Khazaei,~\IEEEmembership{Senior Member,~IEEE}}
}
\IEEEpubidadjcol
\IEEEpubid{\begin{minipage}{\textwidth}\ \\[25pt] \centering
  \color{blue}This work has been submitted to the IEEE for possible publication. Copyright may be transferred without notice, after which this version may no longer be accessible.
\end{minipage}}
\maketitle

\begin{abstract} \normalfont{{\color{black}
The variability of renewable energy generation and the unpredictability of electricity demand create a need for real-time economic dispatch (ED) of assets in microgrids. However, solving numerical optimization problems in real-time can be incredibly challenging. This study proposes using a convolutional neural network (CNN) based on deep learning to address these challenges. Compared to traditional methods, CNN is more efficient, delivers more dependable results, and has a shorter response time when dealing with uncertainties. While CNN has shown promising results, it does not extract explainable knowledge from the data. To address this limitation, a physics-inspired CNN model is developed by incorporating constraints of the ED problem into the CNN training to ensure that the model follows physical laws while fitting the data. The proposed method can significantly accelerate real-time economic dispatch of microgrids without compromising the accuracy of numerical optimization techniques. The effectiveness of the proposed data-driven approach for optimal allocation of microgrid resources in real-time is verified through a comprehensive comparison with conventional numerical optimization approaches. } }   
\end{abstract}
{\color{black}
}

\IEEEpeerreviewmaketitle

\section{Introduction}

Microgrids offer an appealing option for addressing the difficulties posed by aging grid infrastructures and natural disasters on a local scale \cite{wang2022multi}. One of the key practical challenges in microgrid operation is economic dispatch (ED), which involves the optimal allocation of power generation dispatch to meet energy demand \cite{ed1}. The task of finding the optimal microgrid operation can be viewed as an optimization problem that involves multiple soft and hard constraints. These constraints are linked to the specifications of power generation and the need to maintain a balance between energy production and demand, and can vary depending on the available energy resources and cost characteristics of the system\cite{obama2017irreversible,wu2022data}. 
     
Conventional approaches use a model-based numerical optimization techniques to solve the ED problem for microgrids \cite{two-stage, ed2, hybrid,ed3,Stochastic,ed4,mean-tracking,ed5}. For example, in \cite{shuai2018stochastic}, a stochastic optimization algorithm was proposed, which could efficiently solve the uncertainty of the renewable power generation in ED by implementing a new approximate dynamic programming method.
Genetic algorithm is one of the most common methods for optimization problems, which was used in \cite{yeh2020new} to improve the convergence of ED problem and expedite the process of verifying the solutions' feasibility, efficiency, and quality. In addition, with the development of technology, the structure of loads also become diverse, many controllable loads, transferable loads, and heat loads are involved in ED problems. In \cite{hou2020multi}, a multi-objective economic dispatch model considering electrical vehicles and transferable loads is established. Other examples for model-based ED problem include multi-objective optimization approach (theta-dominance-based evolutionary algorithm) with integrated decision-making (fuzzy c-means) \cite{two-stage}, shuffled frog hopping algorithm with particle swarm optimization \cite{hybrid}, frequency-constrained stochastic model to find the optimal generation schedule \cite{Stochastic}, or a mean-tracking model for stochastic economic dispatch ~\cite{mean-tracking}.
     
In comparison to traditional model-based or numerical optimization-based approaches, machine learning-based methods are more efficient, robust, and less reliant on prior knowledge or resource cost characteristics. As the number of constraints increases, directly solving an optimization problem can be computationally complex (e.g., 10-100 times slower than machine learning-based methods \cite{dong2021machine,kalakova2021novel}).  Consequently, machine learning-based techniques have been suggested for solving economic dispatch problems in power systems \cite{dong2021machine,kalakova2021novel,hafeez2020electric, wen2019optimal,gil2019economic,liu2022dynamic,zhou2020combined,ROCCHETTA2019291,du2019intelligent}. For example, in\cite{dong2021machine}, a random-forest regression model was proposed to predict the renewable power generation and local loads and find the optimization operation of a microgrid. 
In \cite{wen2019optimal},  a deep-recurrent neural network with long short-term memory (DRNN-LSTM) was proposed to forecast aggregated load and PV power output during a short-term period and find the optimal load dispatch. The proposed DRNN-LSTM model offered 7.43\% improvement in forecasting and can reduce 8.97\% of the daily cast. In \cite{liu2022dynamic}, an artificial neural network (ANN) model was proposed to deal with the uncertainties brought by renewable generation via applying the nonlinear autoregressive exogenous model (ARIMA). In \cite{zhou2020combined}, the ED problem was modeled  as a markov decision process (MDP) and the authors proposed an improved advanced deep reinforcement learning algorithm to solve the ED problem in combined heat and power (CHP) system. In\cite{ROCCHETTA2019291}, the authors utilized the integration of artificial neural networks and Q-learning algorithms to resolve the optimal management problem of power grid operation and maintenance.

Previous studies have proposed various effective techniques for solving the economic dispatch problem in microgrids. However, these methods are not without limitations. Existing machine learning methods \cite{dong2021machine,kalakova2021novel,hafeez2020electric, wen2019optimal,gil2019economic,liu2022dynamic,zhou2020combined,ROCCHETTA2019291,du2019intelligent,jia2022convopf} heavily rely on collected data and often face difficulties due to operational condition changes. In addition, conventional machine learning techniques have limited generalization capabilities and their convergence rates are difficult to control. Additionally, other machine learning approaches such as reinforcement learning \cite{ROCCHETTA2019291}, decision trees and ARIMA \cite{zhou2020combined} struggle to handle large datasets and require a significant training for accurate predictions, which is time consuming and inefficient. Furthermore, many existing studies only use machine learning methods for forecasting, rather than combining optimization and forecasting to improve performance. In \cite{jia2022convopf}, the authors proposed a CNN model to help solve the optimal flow problem and reduce the prediction running time significantly (350 X speed up). However, the proposed model is purely data-driven and does not include physical rules into the model.  Finally, existing machine learning-based approaches cannot generally adapt to system changes/expansion and require a re-training if a component fails or new components are added. Physics-informed machine learning is an emerging direction in machine learning design that embeds physical laws during the training of a machine learning algorithm and has shown significant improvement in accuracy and adaptation of machine learning techniques to changing situations and uncertainties \cite{karniadakis2021physics}. However, to our best knowledge, no existing study has explored physics-informed machine learning for economic dispatch problem of microgrids. \par 
There is a clear knowledge gap on utilizing machine learning-based approaches for solving the economic dispatch problem in microgrids, which requires further investigation. This paper proposes a physics-inspired machine learning approach via convolutional neural networks (CNN) for solving the ED problem in real time. Due to the time-series nature of economic dispatch problem, two approaches including RNN and CNN could be used. However, CNN has shown promising results and outperformed RNN in many applications, such as data prediction, image classification, and video analysis \cite{harbola2019one}, which has been selected in this work. Furthermore, CNN has multiple convolutional layers, which can improve its performance in extracting and learning features from data compared to deep neural networks (DNNs). The main contributions of this paper are listed as:
     \begin{itemize}    
     \item A 1D-CNN model was developed to solve the economic dispatch problem of a microgrid with renewable generation and conventional generation. Compared with existing deep learning-based approaches, the proposed CNN architecture is provides a more accurate prediction of solution of ED in real-time when intermittencies of renewable generation exists. 
     \item To solve the inability of existing data-driven techniques to account for physical constraints of the system, a physics-inspired CNN approach was developed to incorporate the constraints of the ED problem into the CNN design and increase the accuracy and efficiency of the proposed method.
     \item Incorporating ED constraints in the training of the proposed CNN-based ED architecture significantly reduces the amount of training data required to predict ED solutions in real-time without compromising runtime.
     \item Compared with a conventional ED approach based on numerical optimization, the proposed method is 400 times faster in converging to an optimal solution with an average accuracy more than 98\%.
     \end{itemize}
 Several experiments showcase the accuracy, effectiveness, and real-time computational advantages of the proposed physics-inspired CNN-based ED compared with conventional numerical optimization or deep learning-based techniques.
The rest of the paper is organized as follows: Section II formulates the economic dispatch problem of microgrids  and Section III includes the proposed machine-learning approach. Final results and analysis are included in Section IV and Section V concludes the paper. 

\section{Economic Dispatch of Microgrids}
\begin{figure*}[htp]
\centering
\includegraphics[width=18cm]{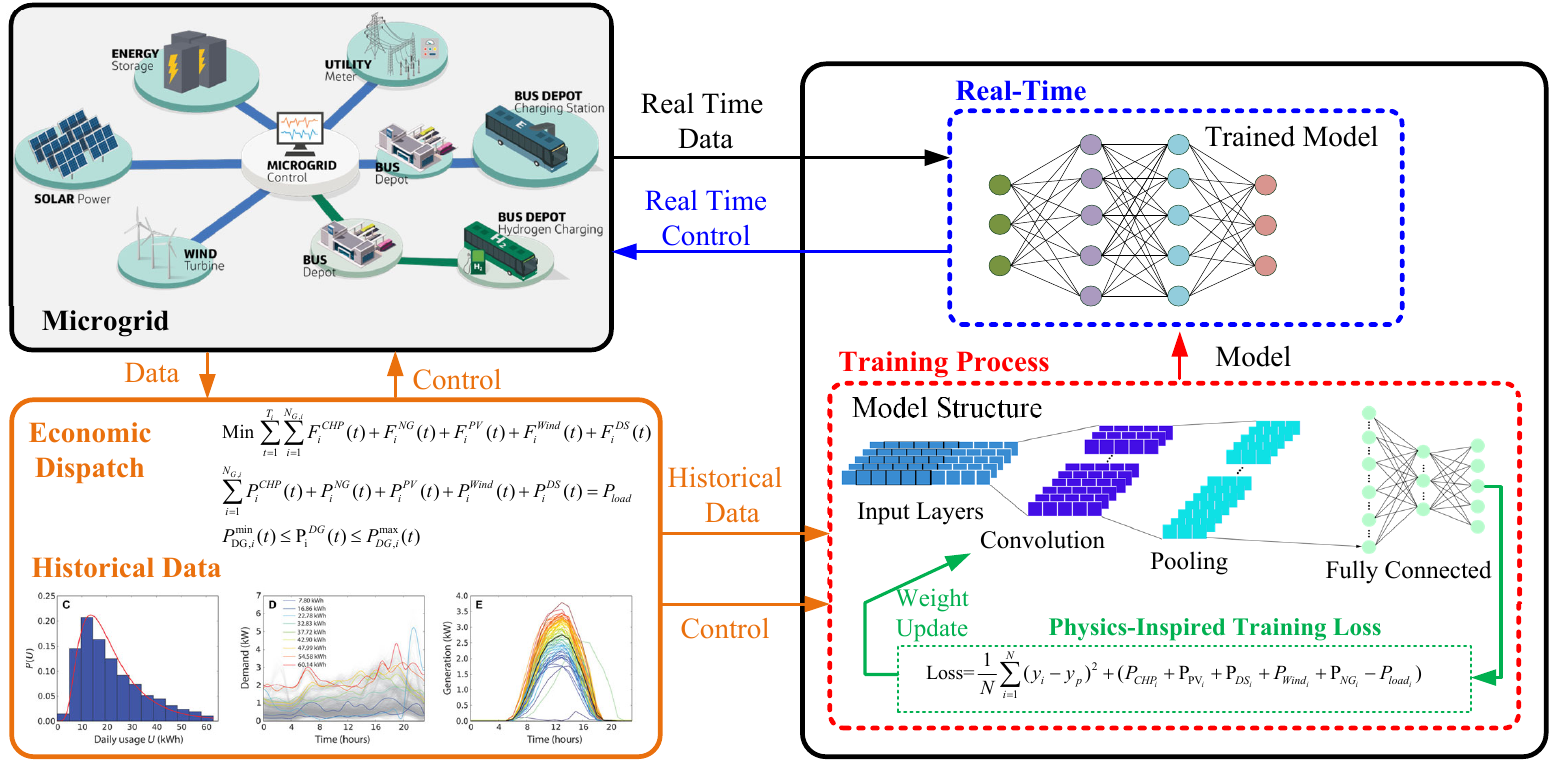}
\caption{Schematic of the proposed PI-CNN based microgrid economic dispatch.}
\label{approach scheme}
\end{figure*}

A microgrid is a community-level power system that includes generation units, energy storage units, and loads. These systems employ various conventional generation resources, such as diesel (DS), natural gas (NG), and combined heat and power (CHP), in addition to renewable resources like solar photovoltaic (PV) and wind turbines (WT), to decrease carbon emissions and minimize power costs. The Economic Dispatch (ED) problem is a typical optimization problem, intending to reduce system costs by rapidly shifting from higher to lower-cost generation before dispatching, all while ensuring a balanced load operation. This problem for a set of diverse generations including CHP, NG, DS, PV, and WT, i.e., $G \in[CHP,\ NG, \ DS, \ PV,\ W]$ can be formulated as a numerical optimization:
\begin{subequations}
\begin{align}
    \mathrm{min} &\sum_{t=1}^{T_i}\sum_{i=1}^{N_{G,i}}c_i(P_{i}^G(t)), \hspace{0.2cm} P_{i}^G\in G \label{obj}\\
s.t. &\quad \sum_{i=1}^{N_{G,i}}P_{i}^G(t)=
P_{load}(t)  \quad G \in \left\{ CG,PV,W \right\} \label{balance}\\
 &U_{G,i}^t P_{G,i}^{\text{min}} \leq P_{i}^G(t) \leq U_{G,i}(t) P_{G,i}^{\text{max}}, \label{eq:gen_limits} \\
&P_{i}^G(t) - P_{i}^G(t-1) \leq R_{\text{up}}^{\text{max}} U_{G,i}(t-1)\label{eq:ramp_up}\\
&P_{i}^G(t-1) - P_{i}^G(t) \leq R_{\text{down}}^{\text{max}} U_{G,i}(t),   \label{eq:ramp_down}  \\
&{P}_{PV,i}^{min}\leq P_{i}^{PV}(t)\leq {P}_{PV,i}^{max} \label{pv}\\
&{P}_{W,i}^{min}\leq P_{i}^{W}(t)\leq {P}_{W,i}^{max} \label{wt}
\end{align}
\end{subequations}
where equation \eqref{obj} is the objective function of the optimization problem to minimize the overall generation cost of generation units $c_i$ within over a time length $T_i$, where $P_{i}^G$ is the generated power (dispatching command) of generation unit $i$. Constrains of this economic dispatch problem are formulated by \eqref{balance}-\eqref{wt}, where the load balance is expressed by \eqref{balance}. Constraints ~\eqref{eq:ramp_up} to \eqref{eq:ramp_down} represent the bounds of generation units and ramp rate conditions with $U_{G,i}$ being the on/off status of the conventional generation, $R_{up}^{max}$ and $R_{down}^{max}$ being the maximum ramp up and ramp down rates  of conventional generators. The cost function $c_i$ for conventional generation is a quadratic function of their generated power, which is formulated by:
\begin{equation}
c_i(P_i^{CHP}(t))=\alpha_i^{CHP}+\beta_i^{CHP}+\gamma_i^{CHP}(P_i^{CHP}(t))^2
\end{equation}
\begin{equation}
c_i(P_i^{NG}(t))=\alpha_i^{NG}+\beta_i^{NG}P_i^{NG}(t)+\gamma_i^{NG}(P_i^{NG}(t))^2
\end{equation}
\begin{equation}
c_i(P_i^{DS}(t))=\alpha_i^{DS}+\beta_i^{DS}P_i^{DS}(t)+\gamma_i^{DS}(P_i^{DS}(t))^2
\end{equation}
where $\alpha,\beta,\gamma$ are the cost function coefficients of generation units. The output of solar generation highly relies on the weather condition, the source of the uncertainty. According to \cite{moazeni2020maximizing} and \cite{hijjo2017pv}, the cost function of PV is formulated as 
\begin{align}
P_i^{PV}(t)&=\rho _{stc}(1+K_T(T_c-T_i))\\
c_i(P_i^{PV}(t))&=K_i^{PV}P_i^{PV}(t)
\end{align}
where $\rho_{stc}=P_{STC}^{PV}\dfrac{I(t)}{I_{STC}}$, $I(t)$ is the solar irradiance at time $t$, $P_{STC}^{PV}$ is the output of the solar generation under standard temperature condition (STC), $I_{STC}$ is the solar radiation at STC, $K_T$ is the temperature coefficient of the solar Photovoltaic panel,  the value of which is -0.0047\cite{hijjo2017pv}, $T_c$ represents the reference temperature and $T_i$ represents the current temperature in degrees. The cost function of solar generation is $c_i(P_i^{PV}(t))$ and $K_i^{PV}$ is the cost coefficient defined as $K_i^{PV}=a_iI_i+G_i$, where $a_i=\dfrac{r_i}{[1-(r_i+1)^{-N}]}$ is the annuitization constant, and $r_i$, $N$, $I_i$, and $G_i$ represent the interest rate, investment lifetime, the investment cost per unit and maintenance cost of PV in \$/kW, respectively.  
The output of the wind generation is associated with wind speed, which is expressed as:
\begin{equation}
P_i^{W}=\left\{
    \begin{array}{lr}
    0\qquad \qquad \qquad \quad \ \text{if} \  v_t \ \textless \ v_{on} \ \text{or} \ v_t \textgreater v_{off}    &  \\
    P_{0}\dfrac{v_t-v_{on}}{v_{0}-v_{on}} \quad \ \  \ \ \ \  \text{if} \  v_{on}\leq v_t \leq v_{0}\\
    P_{0} \qquad \qquad \qquad \ \ \  \text{if} \ v_{0}\leq v_t \leq v_{off} & 
    \end{array}
\right.
\end{equation}
where $P_{0}$ is the generated wind power under a standard wind speed condition ($v_{0}$ ($m/s$)), $v_t$ ($m/s$) is the measured wind speed at time $t$, $v_{on}$ and $v_{off}$ are the excitation speed and cut-off wind speed. 
The cost function of wind turbine is expressed as $c_i(P_i^{Wind}(t))=K_i^{W}P_i^{Wind}(t)$, where $K_i^W$ is the cost coefficient of wind generation, which is calculated similar to the solar energy system as elaborated in detail in \cite{moazeni2021step}.

\section{Physics-inspired Machine Learning-based Approach}

The concept of physics-informed machine learning involves incorporating physical laws or prior knowledge into the learning process to enhance model performance. In this study, we introduce a physics-informed CNN (PI-CNN) model for solving the economic dispatch problem in microgrids. Additionally, we construct a DNN model to provide a comparison and demonstrate the superior performance of the PI-CNN. Fig. \ref{approach scheme} depicts the structure of the proposed approach, where a classical ED problem is utilized and run in an offline fashion to generate data (including inputs to the model and the outputs (result of ED) that are dispatching signals for generation units). The data will be utilized in a PI-CNN to learn the feature. The learned PI-CNN will then be utilized for real-time decision-making. The model structure and training process of DNN and PI-CNN will be introduced in the following sections:

\subsection{Model Structure}
\begin{figure*}[htp]
    \centering
    \includegraphics[width=\textwidth]{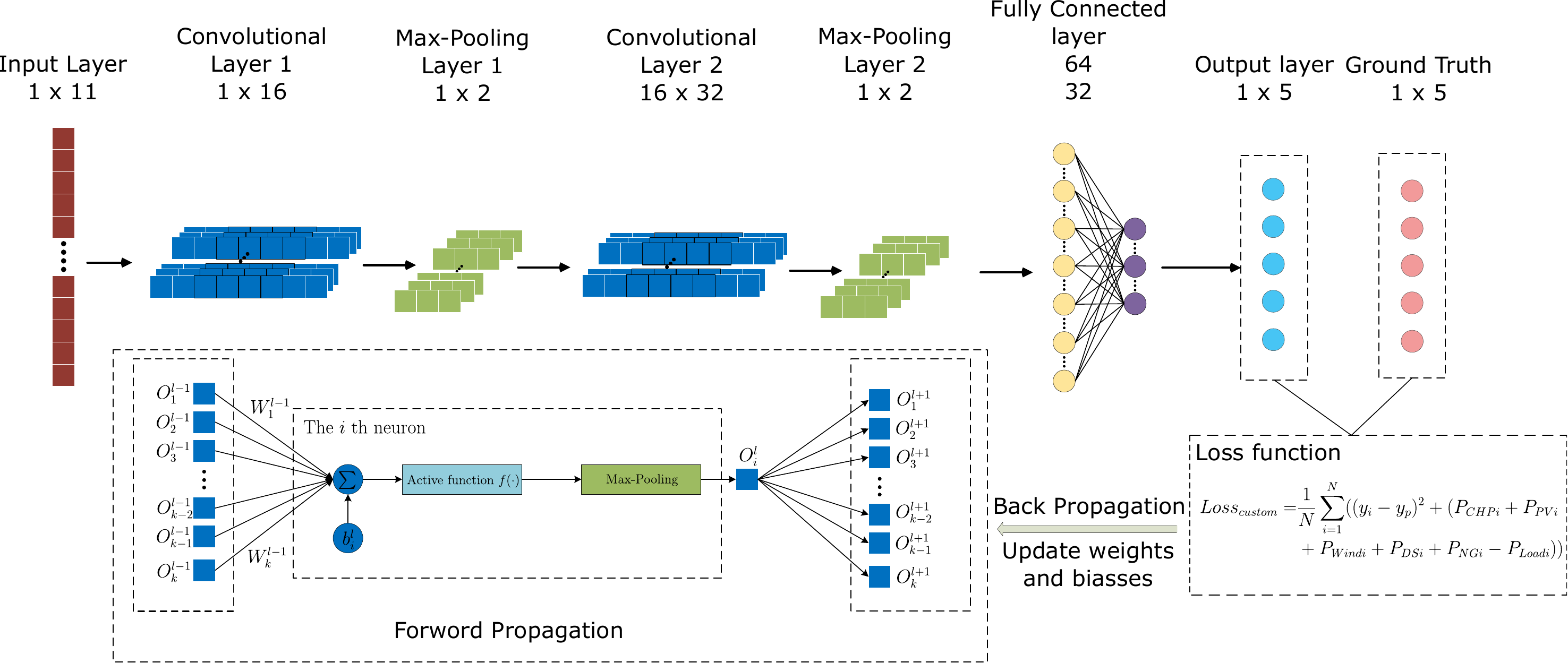}
    \caption{Proposed PI-CNN model structure.}
    \label{cnn structure}
\end{figure*}

The proposed structure of the convolutional neural network is  shown in Fig. \ref{cnn structure}, which is elaborated in the following. 

    
\subsubsection{Convolutional layer}
\begin{figure}[htp]
    \centering
    \includegraphics[width=6cm]{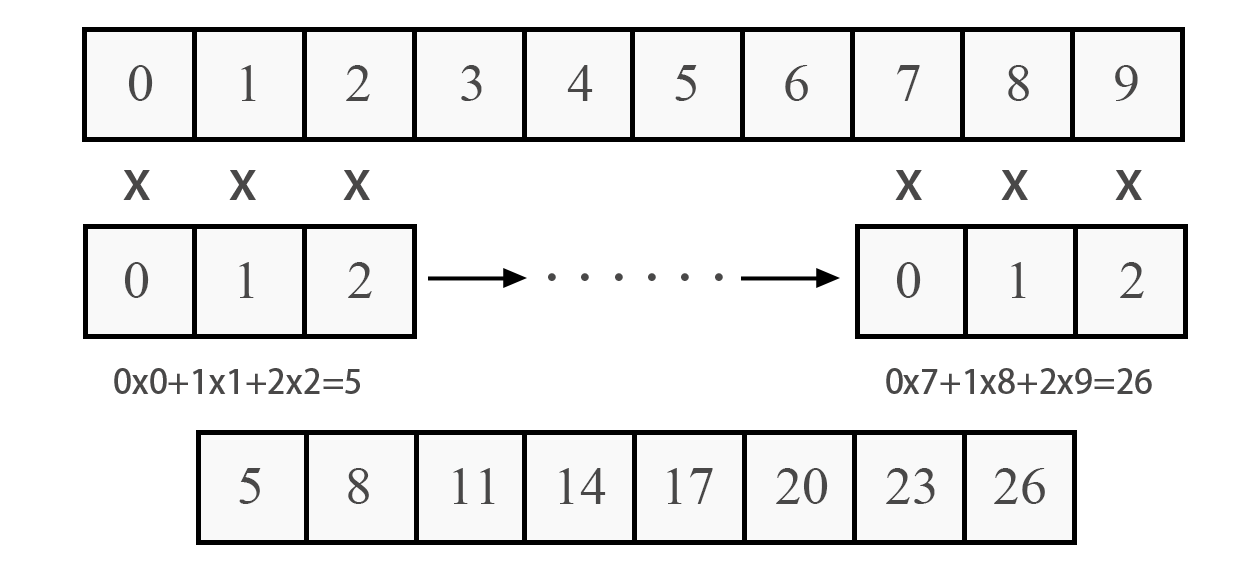}
    \caption{Working process of 1-D Convolution layer.}
    \label{cnn working process}
\end{figure}

1D convolutional layer is used in our approach to extract the feature information from the 1D input data. Conv1D also reduces the huge training parameters's size as well as the training time. The operation of 1D convolutional layer is mathematically represented as:
\begin{equation}
    x_k^l=b_k^l+\sum^{N_{l-1}}_{i=1} w_{ik}^{l-1}\star o_i^{l-1} 
\end{equation}
$x_k^l$ is the input, $b_k^l$ is the $k_{th}$ neuron bias at the $l_{th}$ layer, $o_i^{l-1}$ is the $i_{th}$ neuron output at $l-1_{th}$ layer, $w_{ik}^{l-1}$ is the kernel between the $i_{th}$ neuron at $l-1_{th}$ layer and $k_{th}$ neuron at $l_{th}$ layer, and $\star$ is the convolution operation. Fig. \ref{cnn working process} shows the working process.

 \subsubsection{Activation function} In our approach, we choose the ReLU as the activation function. The equation representing the ReLU function is expressed by $f(x)=max(0,x) $, which means it returns 0 when the input is negative and keeps the original value when the input is positive. Because of this property, ReLU is considered as the most popular activation function. Comparing with the linear activation function, ReLU doesn't have the vanishing gradient problem and it is more efficient\cite{krizhevsky2017imagenet}.
    
\subsubsection{MAX-Pooling layer} 
\begin{figure}
    \centering
    \includegraphics[width=8.5cm]{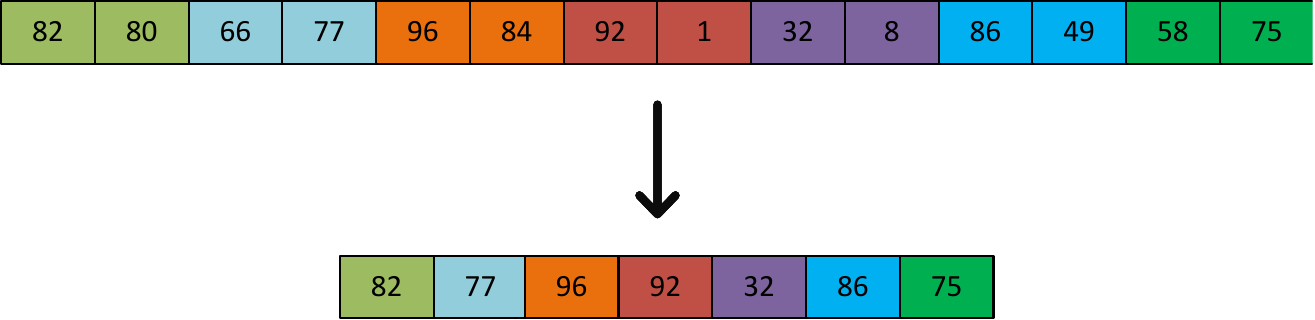}
    \caption{Working process of max-pooling layer.}
    \label{maxpooling}
\end{figure}

The max-pooling layers are used to avoid overfitting problems and improve the model's robustness by reducing the data dimensions and removing redundant information \cite{afrasiabi2019probabilistic}. 
The working process is shown in Fig.\ref{maxpooling}

\subsubsection{Fully connected layer} Fully connected layers with different locations (i.e., fully connected layers in the input or output layers) have different functions. The fully connected input layer takes the output of the previous layers, flattens them and turns them into a single vector that can be an input for the next process. The fully connected layers in the middle location take the inputs from the feature analysis and apply weights to predict the value. Fully connected output layer gives the final predicted values for each output. 

\begin{figure}
    \centering
    \includegraphics[width=5cm]{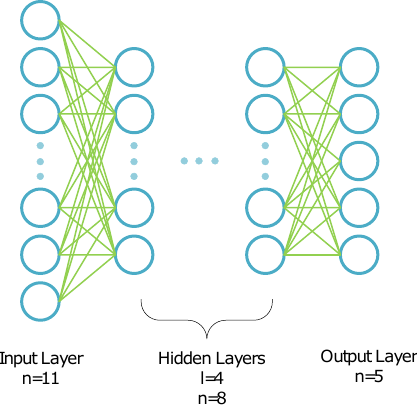}
    \caption{DNN model structure.}
    \label{dnn structure}
\end{figure}

The structure of DNN is shown in Fig. \ref{dnn structure}. There are 6 layers in the proposed DNN model, including 1 input layer, 1 output layer, and 4 hidden layers. Each hidden layer has 8 neurons. The optimizer and loss function are similar to those used in the CNN. 
\subsection{Loss Function Design}
The loss function is the function to determine the distance between the model outputs and the ground truth. In the training process, the loss will reduce with the weights and bias updated in the back propagation process. Hence, to achieve the best performance, it is necessary to design targeted loss functions based on different tasks. Mostly, cross-entropy loss function is used for classification tasks and mean square error (MSE) loss function is utilized for regression tasks.
Here, the MSE loss is chosen to be used for the ED problem. For the physics-inspired (PI) part, a constraint is added to the loss function, which acts as the prior knowledge of the loss function. It can help the model learn how to fit the observed data while forcing the prediction to fulfilled the given physical constraints. This kind of constraint can help one integrate more physical laws based on the constraints of the model into the formulation.
\subsubsection{MSE Loss} The MSE loss is formulated as: 
\begin{equation}
    \mathcal{L}oss_{MSE}=\dfrac{1}{N}||y_i-y_p||^2_2
\end{equation}
where N is the batch size, $y_i$ is the ground truth and $y_p$ is the predicted value. 
\subsubsection{Physics-inspired Loss}


The goal of PI-loss is to aid the model learning the robust relationship between load and different generators described in \eqref{balance} to improve the model performance. We design the physics-informed loss function based on the MSE loss function and the power balance constraint (PBC), $ \sum_{i=1}^{N_{G,i}}P_{i}^G(t)=P_{load}(t) \quad G \in \left\{ CG,PV,W \right\}$. We add the power balance constrain $P_{i}^{CHP}+P_{i}^{PV}+P_{i}^W+P_{i}^{NG}-P_{Load}=\sum_{i=1}^{N_{G,i}}P_{i}^G -
P_{load}=0$ to the original MSE Loss function. The physics-informed loss function is defined as: 
\begin{align}\label{eq.zocl}
\mathcal{L}oss_{PI}&=\mathcal{L}oss_{MSE}+\mathcal{L}oss_{PBC}\notag \\
&=\dfrac{1}{N}\lVert y_i-y_p\lVert^2_2+\lambda_1 \lVert \sum_{i=1}^{N_{G,i}}P_{G,i} -
P_{load} ||^2_2
\end{align}
where $\lambda_1$ is a hyperparameter to penalize the solutions that violate the power balance. In addition, to ensure the solution of the proposed PI-based algorithm (i.e., generation dispatch) is always within the bounds and the ramp rate conditions are not violated, the constraints \eqref{eq:gen_limits}-\eqref{wt} will also need to be added to the loss function. 
To add the constraints to the loss function, we need to first convert the inequality constraints to equality constraints. Utilizing concepts from convex optimization theories, we first separate each constraint in \eqref{eq:gen_limits}-\eqref{wt}  into two constraints and add a slack variable to each inequality constraint to convert it to an equality constraint suitable to be embedded in the loss function of the machine learning. The resulting constraints are written as
\begin{align}
    &s_1 - P_{i}^G(t) =-U_{G,i}(t) P_{G,i}^{\text{min}}, \label{eq11} \\
       &s_2+ P_{i}^G(t) = U_{G,i}(t) P_{G,i}^{\text{max}},  \\
&s_3+ P_{i}^G(t) = P_{G,i}(t-1) +R_{\text{up}}^{\text{max}} U_{G,i}(t-1)\\
& s_4- P_{i}^G(t) =R_{\text{down}}^{\text{max}} U_{G,i}^t-P_{i}^G(t-1),    \\
&s_5- P_{i}^{PV}(t)=-{P}_{PV,i}^{min} \\
&s_6+P_{i}^{PV}(t)= {P}_{PV,i}^{max} \\
&s_7- P_{i}^W(t)=-{P}_{W,i}^{min} \\
&s_8+ P_{i}^W(t)= {P}_{W,i}^{max} \label{eq18}
\end{align}
where $s_1-s_6$ are six slack variables defined to convert the inequality constraints to equality constraints. The new equality constraints in \eqref{eq11}-\eqref{eq18} can now be added to the loss function of the machine learning as
\begin{align}\label{eq19}
\mathcal{L}oss_{PI}&=\mathcal{L}oss_{MSE}+\mathcal{L}oss_{PBC}+\mathcal{L}oss_{Con} \\
\mathcal{L}oss_{Con}&=\lambda_2\lVert s_1-P_i^G(t)+U_{G,i}(t)P_{G,i}^{min}\lVert^2_2+ \notag\\
&\lambda_3\lVert s_3+ P_{i}^G(t) - P_{G,i}(t-1) -R_{\text{up}}^{\text{max}} U_{G,i}(t-1)\lVert^2_2+ \notag\\
&\lambda_4\lVert s_4- P_{i}^G(t) -R_{\text{down}}^{\text{max}} U_{G,i}^t
+P_{i}^G(t-1)\lVert^2_2+ \notag\\
&\lambda_5\lVert s_5- P_{i}^{PV}(t)+{P}_{PV,i}^{min}\lVert^2_2+ \notag\\
&\lambda_6\lVert s_6+P_{i}^{PV}(t)- {P}_{PV,i}^{max}\lVert^2_2+ \notag\\
&\lambda_7\lVert s_7- P_{i}^W(t)+{P}_{W,i}^{min}\lVert^2_2+ \notag\\
&\lambda_8\lVert s_8+ P_{i}^W(t)- {P}_{W,i}^{max}\lVert^2_2
\end{align}
The combined loss function presented in \eqref{eq19} will ensure the constraints of the original economic dispatch function are embedded within the training of machine learning. It is also noted that $\lambda_2$-$\lambda_8$ are new hyperparameters designed to penalize the violation of economic dispatch constraints and are tuned manually during the training process. 
\subsection{PI-CNN Working Process}
Forward propagation and back propagation are two parts of CNN's working process. The former involves sequentially calculating and preserving intermediate variables from the input layer to the output layer, as discussed above. Back propagation is a technique used for computing the gradient of CNN for updating the CNN parameters. Based on the chain rule from calculus, this method traverses the network in reverse order, starting from the output layer and moving towards the input layer. Its main purpose is to minimize the difference between predicted and actual output in order to update the parameters of the network \cite{kiranyaz20211d}. The process from the output layer to the last convolutional layer can be mathematically presented as: 
\begin{equation}
    \frac{\partial \mathcal{L}}{\partial o_{k}^{l}}=\sum_{i=1}^{N_{l+1}}\frac{\partial \mathcal{L}}{\partial x_i^{l+1}}\frac{\partial x_i^{l+1}}{\partial o_k^{l}}=\sum_{i=1}^{N_{l+1}}\Delta_i^{l+1}w^l_{ki}
\end{equation}
where $\mathcal{L}$ is the loss value. When the process proceeds to the next convolution layer, the delta error is: 
\begin{equation}
\Delta_k^l=\frac{\partial \mathcal{L}}{\partial y_k^l}\frac{\partial y_k^l}{\partial x_k^l}=\frac{\partial \mathcal{L}}{\partial uo_k^l}\frac{\partial uo_k^l}{\partial y_k^l}f'(x_k^l)
\end{equation}
where $uo_k^l$ is the zero-order up sampling and $f(\cdot)$ is the activation function. 
The delta error through back propagation can be expressed as: 
\begin{equation}
    \Delta o_k^l=\sum_{i=1}^{N_{l+1}}\Delta_i^{l+1}\star R(w_{ki}^l)
\end{equation}
where $R(\cdot)$ is the reverse function. Based on this, the sensitivities of weight and bias can be calculated as: 
\begin{equation}
\frac{\partial \mathcal{L}}{\partial w_{ik}^l}=o_k^l \star \Delta_i^{l+1}
\end{equation}
\begin{equation}
\frac{\partial \mathcal{L}}{\partial b_{k}^l}=\sum_n\Delta_k^l(n)
\end{equation}
Then, the weight and bias can be updated by:
\begin{equation}
    w_{ik}^{l-1}(t+1)=w_{ik}^{l-1}(t)-\varepsilon \frac{\partial \mathcal{L}}{\partial w_{ik}^{l-1}}
\end{equation}
\begin{equation}
    b_{k}^{l}(t+1)=b_{k}^{l}(t)-\varepsilon \frac{\partial \mathcal{L}}{\partial b_{k}^{l}}
\end{equation}
\section{Case Studies}
To verify the performance of the proposed CNN design, several case studies were conducted that include: i) comparison between the performance of a regular  CNN and conventional DNN, ii) comparison between the performance of a regular CNN and physics inspired CNN, iii) the impact of the size of training data on physics-inspired CNN, iv) runtime comparison, and v) performance/feature comparison with the state-of-the-art machine learning-based microgrid ED. The case studies are conducted using torch library with PyTorch in python on a workstation with Intel Core i7-10870H CPU@2.20GHZ, 16G RAM and NVIDIA GeForce RTX-3070 GPU. 
\begin{table*}[]
\centering
\caption{Numerical evaluation  criterion for DNN and CNN}
\label{tab:my-table}
\begin{tabular}{|cccccccccc}
\hline
\multicolumn{10}{c}{Numerical evaluation  criterion} \\ \hline
\multicolumn{1}{c}{} & \multicolumn{3}{c}{MSE} & \multicolumn{3}{c}{$R^2$} & \multicolumn{3}{c}{MAE} \\ \hline
\multicolumn{1}{c}{Generation} & \multicolumn{1}{c}{CNN} & \multicolumn{1}{c}{DNN-150} & \multicolumn{1}{c}{DNN-500} & \multicolumn{1}{c}{CNN} & \multicolumn{1}{c}{DNN-150} & \multicolumn{1}{c}{DNN-500} & \multicolumn{1}{c}{CNN} & \multicolumn{1}{c}{DNN-150} & DNN-500 \\ \hline
\multicolumn{1}{c}{CHP} & \multicolumn{1}{c}{0.1517} & \multicolumn{1}{c}{0.1609} & \multicolumn{1}{c}{0.1112} & \multicolumn{1}{c}{0.9588} & \multicolumn{1}{c}{0.9563} & \multicolumn{1}{c}{0.9698} & \multicolumn{1}{c}{0.2185} & \multicolumn{1}{c}{0.2393} & 0.2327 \\ \hline
\multicolumn{1}{c}{NG} & \multicolumn{1}{c}{0.0333} & \multicolumn{1}{c}{0.2836} & \multicolumn{1}{c}{0.0390} & \multicolumn{1}{c}{0.9908} & \multicolumn{1}{c}{0.9216} & \multicolumn{1}{c}{0.9892} & \multicolumn{1}{c}{0.0409} & \multicolumn{1}{c}{0.3914} & 0.1146 \\ \hline
\multicolumn{1}{c}{DS} & \multicolumn{1}{c}{0.0133} & \multicolumn{1}{c}{0.4404} & \multicolumn{1}{c}{0.0503} & \multicolumn{1}{c}{0.9961} & \multicolumn{1}{c}{0.8719} & \multicolumn{1}{c}{0.9854} & \multicolumn{1}{c}{0.1397} & \multicolumn{1}{c}{0.4693} & 0.1616 \\ \hline
\multicolumn{1}{c}{Wind} & \multicolumn{1}{c}{1.3402} & \multicolumn{1}{c}{3.2092} & \multicolumn{1}{c}{3.1659} & \multicolumn{1}{c}{0.9942} & \multicolumn{1}{c}{0.9862} & \multicolumn{1}{c}{0.9885} & \multicolumn{1}{c}{0.6182} & \multicolumn{1}{c}{1.1664} & 1.0857 \\ \hline
\multicolumn{1}{c}{Sol} & \multicolumn{1}{c}{1.4872} & \multicolumn{1}{c}{2.4672} & \multicolumn{1}{c}{2.4479} & \multicolumn{1}{c}{0.9973} & \multicolumn{1}{c}{0.9955} & \multicolumn{1}{c}{0.9968} & \multicolumn{1}{c}{0.5643} & \multicolumn{1}{c}{0.8716} & 0.8230 \\ \hline
\end{tabular}
\end{table*}

\subsection{Datasets}
\begin{figure}
    \centering
    \includegraphics[width=3.3in]{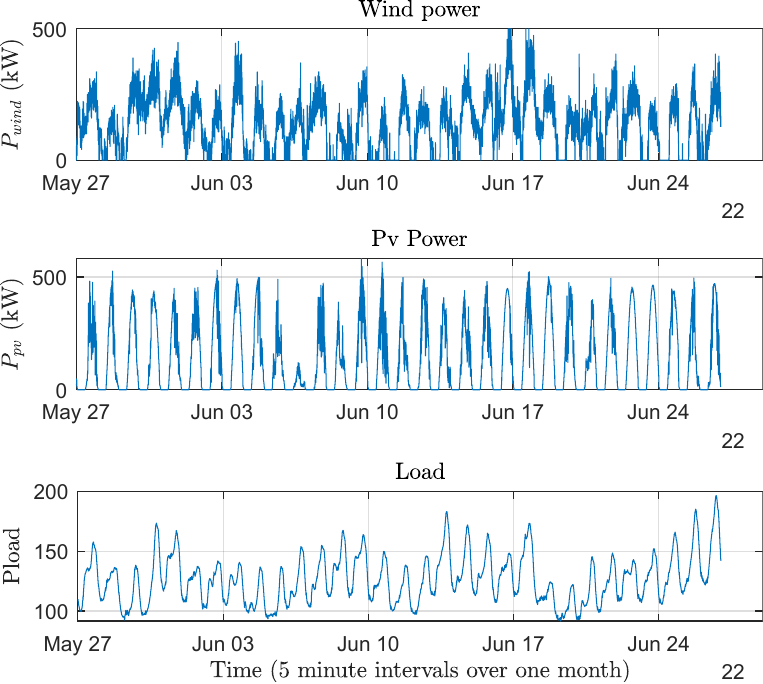}
    \caption{{Input data of $P_{wind}$,\, $P_{Pv}$,\, $Load$}.}
    \label{input data of pwind}
\end{figure}

The numerical solution of the economic dispatch \eqref{obj}-\eqref{wt} is generated by the OPTI Toolbox in MATLAB with "BONMIN" solver \cite{currie2012opti}. The dataset contains 8928 samples, each with 11 features (inputs) and 5 targets (generation unit dispatch), which represents the solution and constraints for one month of data at a 5-minute resolution interval. The input data for wind power, solar PV power and load data for one month is shown in Fig.\,\ref{input data of pwind}. 
Additional features in the dataset include the minimum and maximum limits of solar power and wind power, as well as the upper and lower bounds for CHP, natural gas power plant, and diesel power plant. The microgrid parameters and energy source limits were sourced from \cite{moazeni2021step}.
The targets in this dataset refer to the power required from each energy source to fulfill the demand, which includes the power generated from solar, wind, CHP, natural gas, and diesel. The upper bound of the renewable generations is set as the maximum value of the prediction shown in Fig. \ref{input data of pwind}.
\subsection{Evaluation Metrics}
To accurately and comprehensively evaluate our algorithm, we used three different evaluation metrics:
\subsubsection{$R^2$}
Coefficient of determination ($R^2$) is a statistical measure that represents the proportion of the variance for a dependent variable that is explained by an independent variable or variables in a regression model, which is defined as:  
\begin{equation}
    R^2=1-\frac{\sum_{i=1}^N{(y_i-y_{pi})^2}}{\sum_{i=1}^N{(y_i-y_m)^2}}
\end{equation}
where $y_{pi}$ is the predicted value, $y_i$ is the ground truth, and $y_m$ is the mean value of the ground truth. The closer $R^2$ is to 1, the higher the goodness of fit of the model.  
\subsubsection{MAE}
Mean absolute error (MAE) is a measure of absolute errors between paired observations. The smaller the value is, the better the result is. 
\begin{equation}
    MAE=\frac{1}{N}||y_i-y_{pi}||
\end{equation}
\subsubsection{MSE}
Mean squared error (MSE) measures the average squared difference between the predicted values and the ground truth. The smaller the value is, the better the performance is. 
\begin{equation}
    MSE=\frac{1}{N}||y_i-y_{pi}||^2_2
\end{equation}

\subsection{Data-Prepossessing}
Due to the considerable variation in the training data range, data normalization is applied before training. In this context, Min-max normalization is employed, which is formulated as:
\begin{equation}
X_\sigma  =\frac{X_i-X_{min}}{X_{max-X_{min}}} 
\end{equation}
where $X_{min}$,$X_{max}$ are the minimum and maximum of training data. 
\subsection{Results}
This subsection aims to analyze the effectiveness of the proposed design in different case studies. The baseline setting for all case studies use 80\% of data as the training data and 20\% of the data as the testing data. All above performance metrics are used to provide a comprehensive performance evaluation. 
\begin{figure}
    \centering
    \includegraphics[width=9cm]{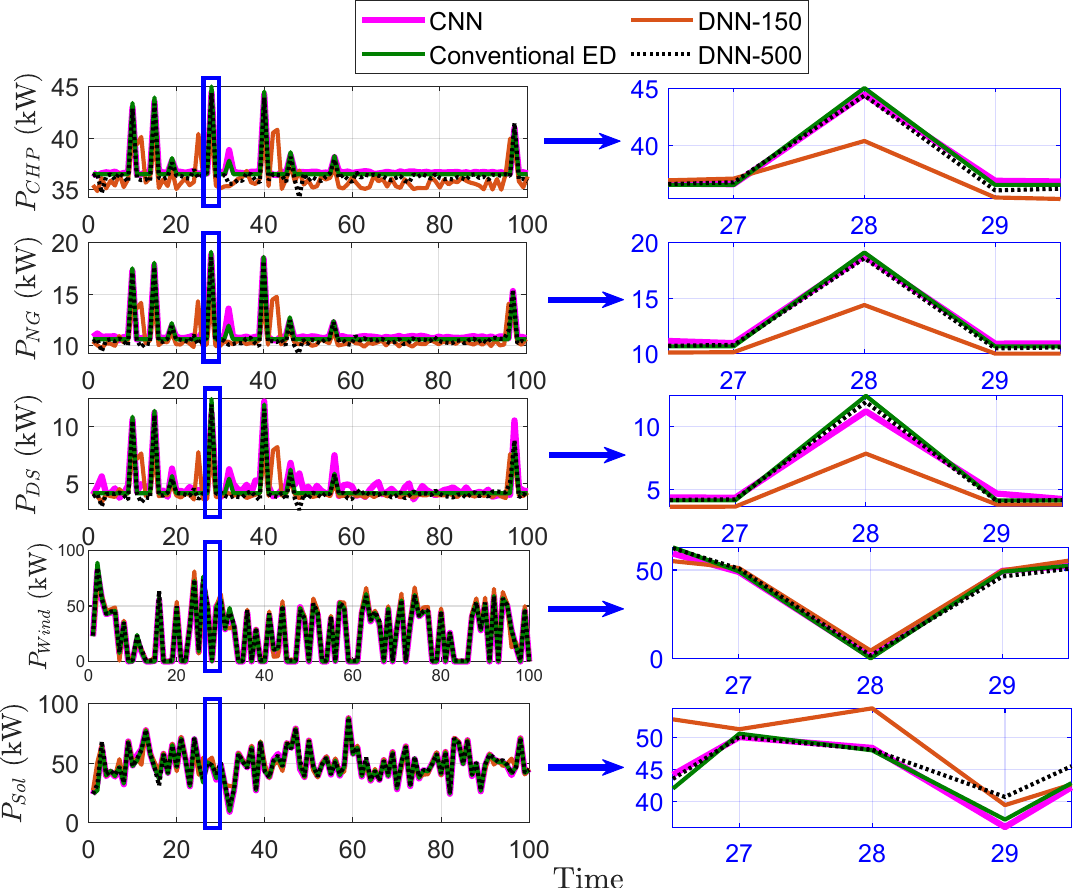}
    \caption {Comparing the ED results of DNN and CNN.}
    \label{results comparion between DNN and CNN}
\end{figure}
\subsubsection{Comparison between the CNN and DNN}
The aim of this study is to compare the predictive performance of CNN and DNN models and demonstrate the feasibility and efficiency of CNN. Both the CNN and DNN-150 models were trained with 150 epochs, while DNN-500 had the same model structure as DNN-150 but was trained with 500 epochs. The results are presented in Fig.\,\ref{results comparion between DNN and CNN}, and the numerical evaluation results are illustrated in Table. \ref{tab:my-table}  

As can be observed in Fig.\,\ref{results comparion between DNN and CNN}, comparing the prediction results of DNN-150 and the CNN, it is concluded that CNN can fit the ground truth better. The numerical results in Table. \ref{tab:my-table} indicate that it is hard for DNN-150 to predict all the outputs from different generation accurately, while the CNN trained at the same number of epochs has a much better performance. It is also observed that by increasing the number of training epochs, the performance of DNN will improve greatly. However, the MAE and MSE values of DNN-500 are still at least 6\% less than the corresponding values for the CNN approach, which showcases the feasibility and efficiency of the proposed CNN-based economic dispatch in accurately predicting the real-time dispatch of generation units.

\subsubsection{Comparison between Conventional CNN and Physics-inspired CNN}
\begin{figure}
    \centering
    \includegraphics[width=3.3in]{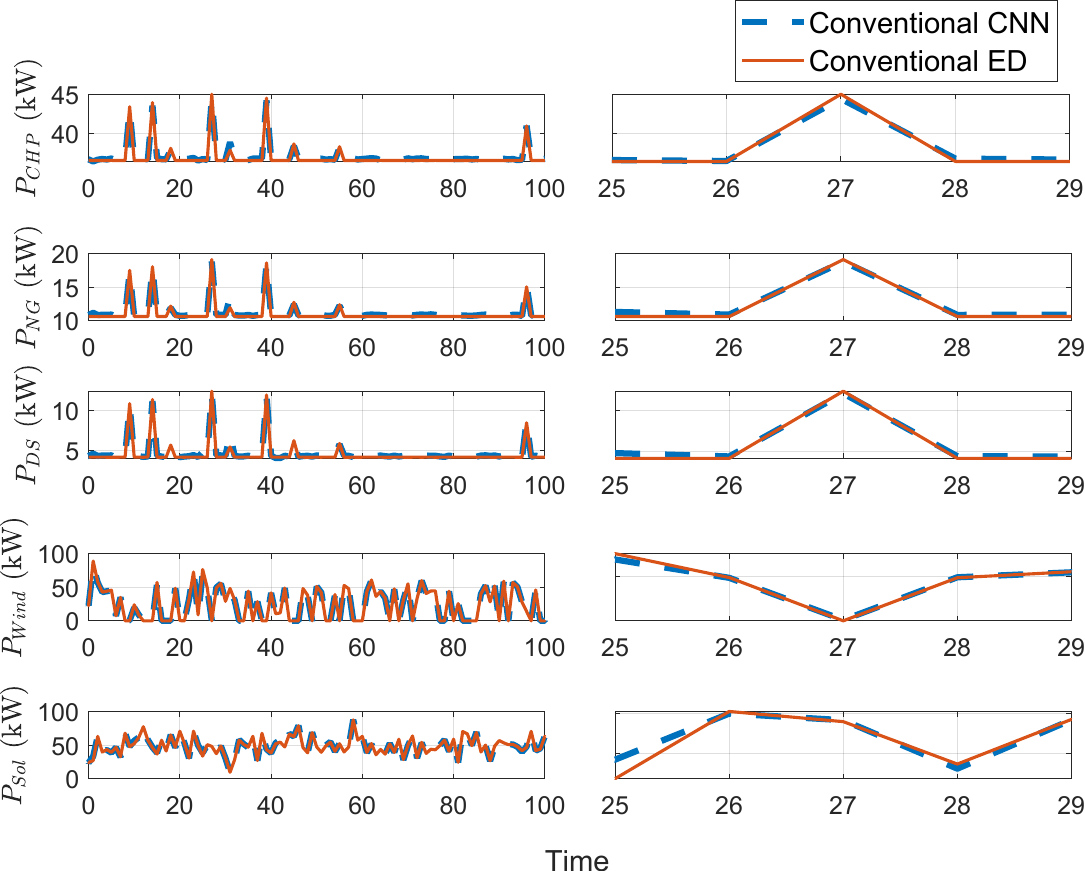}
    \caption{The ED results of conventional CNN.}
    \label{Prediction results from Conventional CNN}
\end{figure}

\begin{figure}
 \centering
\includegraphics[width=\columnwidth]{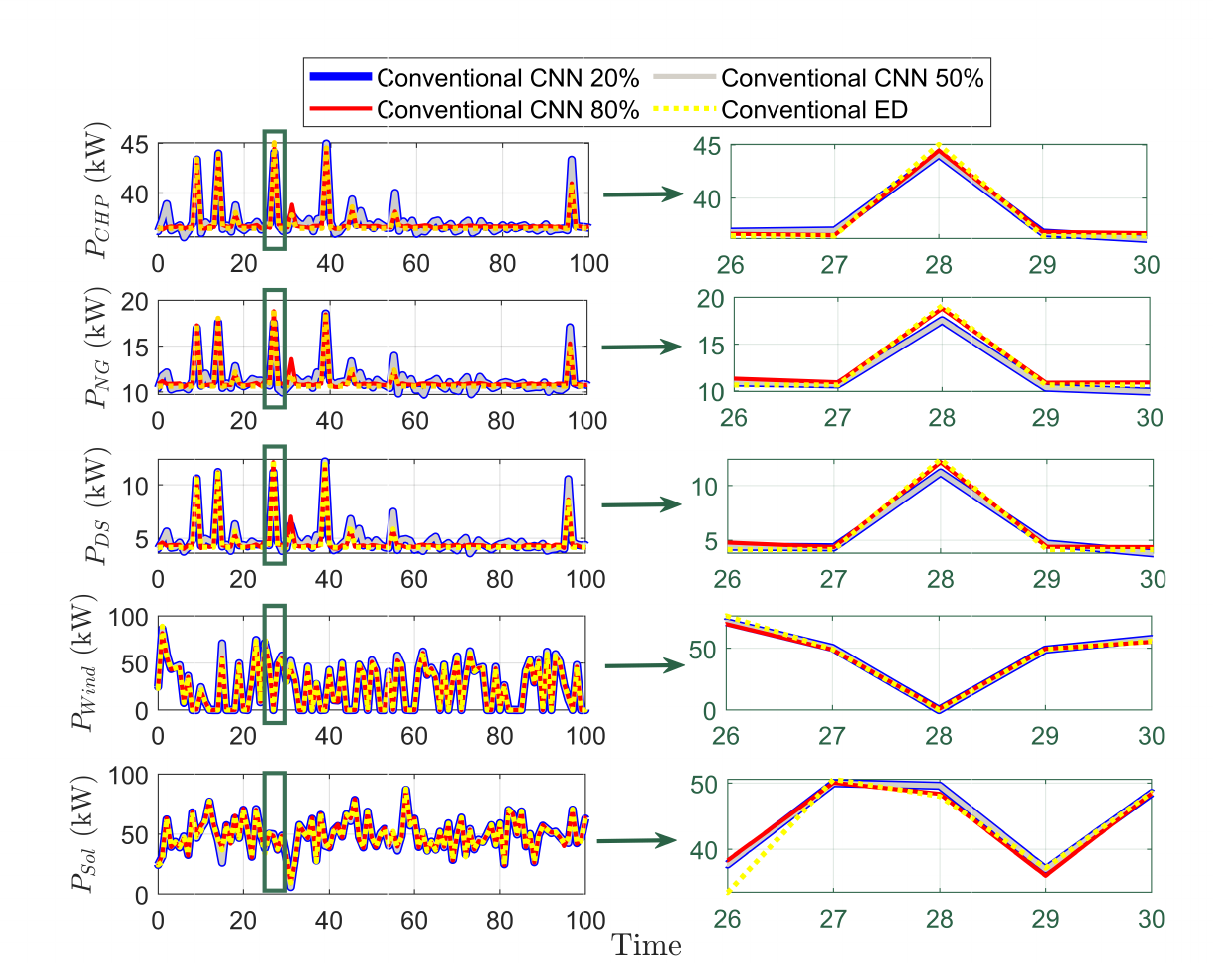}
\caption{Conventional CNN performance for different size of training data: 20\%, 50\% and 80\%.}
\label{Conventional CNN performance comparison}
\end{figure}

\begin{figure}
\centering
\includegraphics[width=\columnwidth]{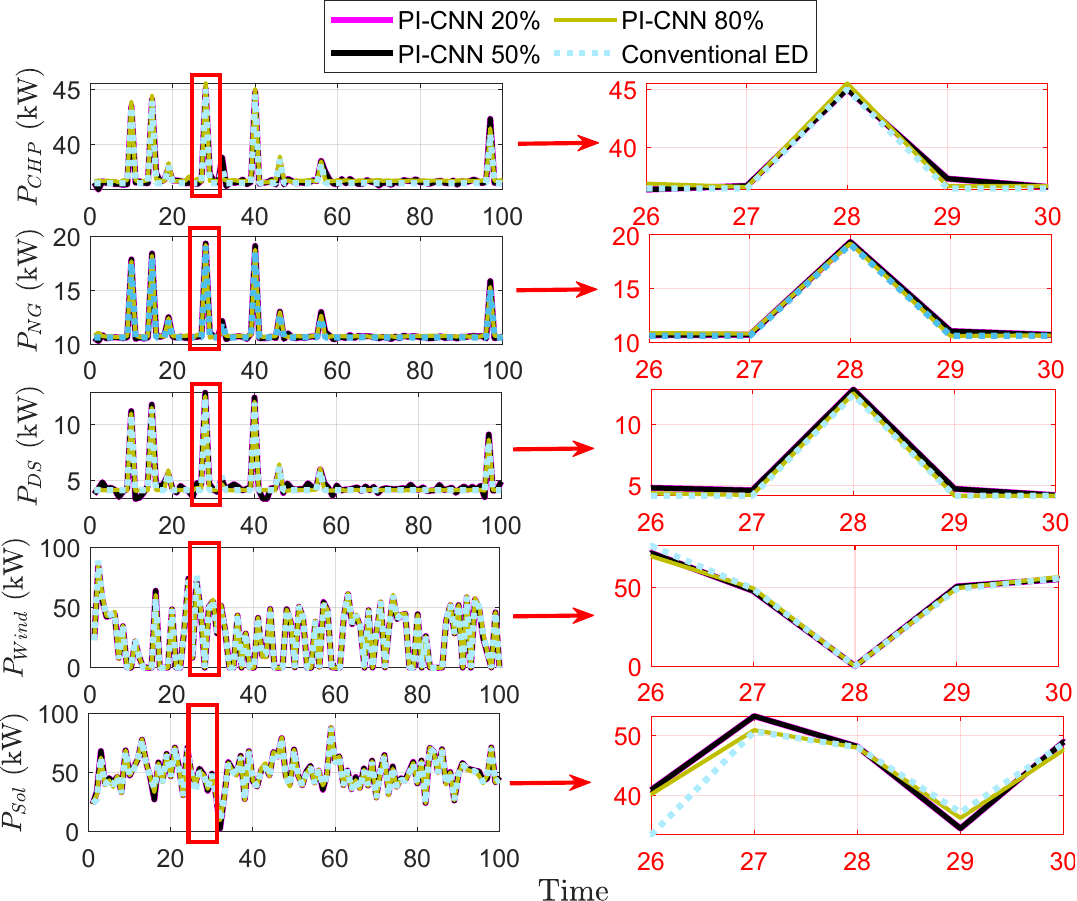}
\caption{PI-CNN performance for different size of training data: 20\%, 50\% and 80\%.}
\label{PI-CNN performance comparison }
\end{figure}

Although the conventional CNN showed a high performance compared with the DNN in the last case study, it still has some limitations. More importantly, the prediction accuracy will decrease with the reduction of the training data size. The physics law embedded in the PI-CNN can make up those limitations. This case study aims to compare the performance of the CNN and PI-CNN, and show the advantages of PI-CNN. 
\begin{figure}
    \centering
    \includegraphics[width=3.5in]{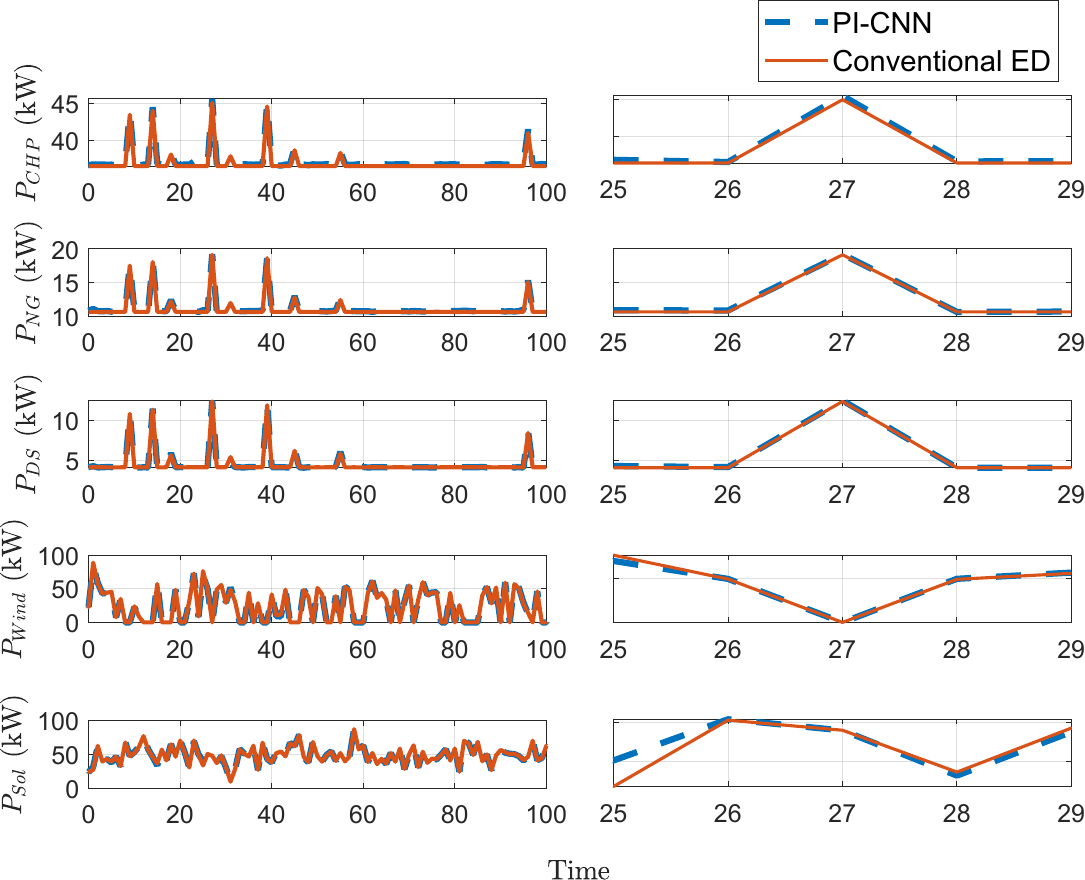}
    \caption{The ED results of physics-inspired CNN.}
    \label{Prediction results from PI-CNN}
\end{figure}
Fig.\,\ref{Prediction results from PI-CNN} and Fig.\,\ref{Prediction results from Conventional CNN}  represent the prediction results, where the left column is the original results and the right column is the zoomed in version of the results. The numerical evaluation criterion is listed in  Table. \ref{tab:my-table1}. From the numerical evaluation of performance metrics, it is observed that under the base training condition (80\% training data and 150 epochs), the PI-CNN has a much lower MSE for CHP, NG, DS, and solar and has a better performance than conventional CNN. In addition, Fig.\,\ref{Conventional CNN performance comparison} and Fig.\,\ref{PI-CNN performance comparison } show the prediction results of conventional CNN and PI-CNN with different size of training data. The prediction accuracy reduces with reduction of the training data for both of those two models, while the PI-CNN shows the higher stability and efficiency when the size of the training data is smaller. Fig.\,\ref{MSE MAE value comparison } shows the MSE, $R^2$ and MAE values with different sizes of training data. From those three figures, both conventional CNN and PI-CNN have a stable high-quality performance when the size of training data is decreasing, even with only 20\% training data, which proves the stability of the proposed CNN model. Furthermore, in most cases, PI-CNN has a better performance than the conventional CNN as is evident from the $R^2$ values, especially for the NG, with its $R^2$ value surpassing conventional CNN by up to 0.67\%, and at least 6\% lower in MSE and MAE value under the same training condition. 

\begin{table}[]
\caption{Performance metrics of PI-CNN and CNN}
\label{tab:my-table1}
\begin{tabular}{lllllll}
\hline
\multicolumn{7}{c}{Numerical evaluation  criterion} \\ \hline
\multicolumn{1}{l}{} & \multicolumn{2}{c}{MSE} & \multicolumn{2}{c}{$R^2$} & \multicolumn{2}{c}{MAE} \\ \hline
\multicolumn{1}{l}{Gen.} & \multicolumn{1}{l}{PI-CNN} & \multicolumn{1}{l}{CNN} & \multicolumn{1}{l}{PI-CNN} & \multicolumn{1}{l}{CNN} & \multicolumn{1}{l}{PI-CNN} & CNN \\ \hline
\multicolumn{1}{l}{CHP} & \multicolumn{1}{l}{0.0379} & \multicolumn{1}{l}{0.1517} & \multicolumn{1}{l}{0.9897} & \multicolumn{1}{l}{0.9855} & \multicolumn{1}{l}{0.1581} & 0.2185 \\ \hline
\multicolumn{1}{l}{NG} & \multicolumn{1}{l}{0.0092} & \multicolumn{1}{l}{0.0333} & \multicolumn{1}{l}{0.9975} & \multicolumn{1}{l}{0.9908} & \multicolumn{1}{l}{0.0635} & 0.0049 \\ \hline
\multicolumn{1}{l}{DS} & \multicolumn{1}{l}{0.0093} & \multicolumn{1}{l}{0.0133} & \multicolumn{1}{l}{0.9973} & \multicolumn{1}{l}{0.9961} & \multicolumn{1}{l}{0.0851} & 0.1379 \\ \hline
\multicolumn{1}{l}{Wind} & \multicolumn{1}{l}{1.4370} & \multicolumn{1}{l}{1.3402} & \multicolumn{1}{l}{0.9938} & \multicolumn{1}{l}{0.9942} & \multicolumn{1}{l}{0.4978} & 0.6182 \\ \hline
\multicolumn{1}{l}{Sol} & \multicolumn{1}{l}{1.4122} & \multicolumn{1}{l}{1.4872} & \multicolumn{1}{l}{0.9974} & \multicolumn{1}{l}{0.9973} & \multicolumn{1}{l}{0.4588} & 0.5643 \\ \hline
\end{tabular}
\end{table}

\begin{figure}[htbp]
\centering
\subfloat[MAE Comparison]{
    \includegraphics[scale=0.5]{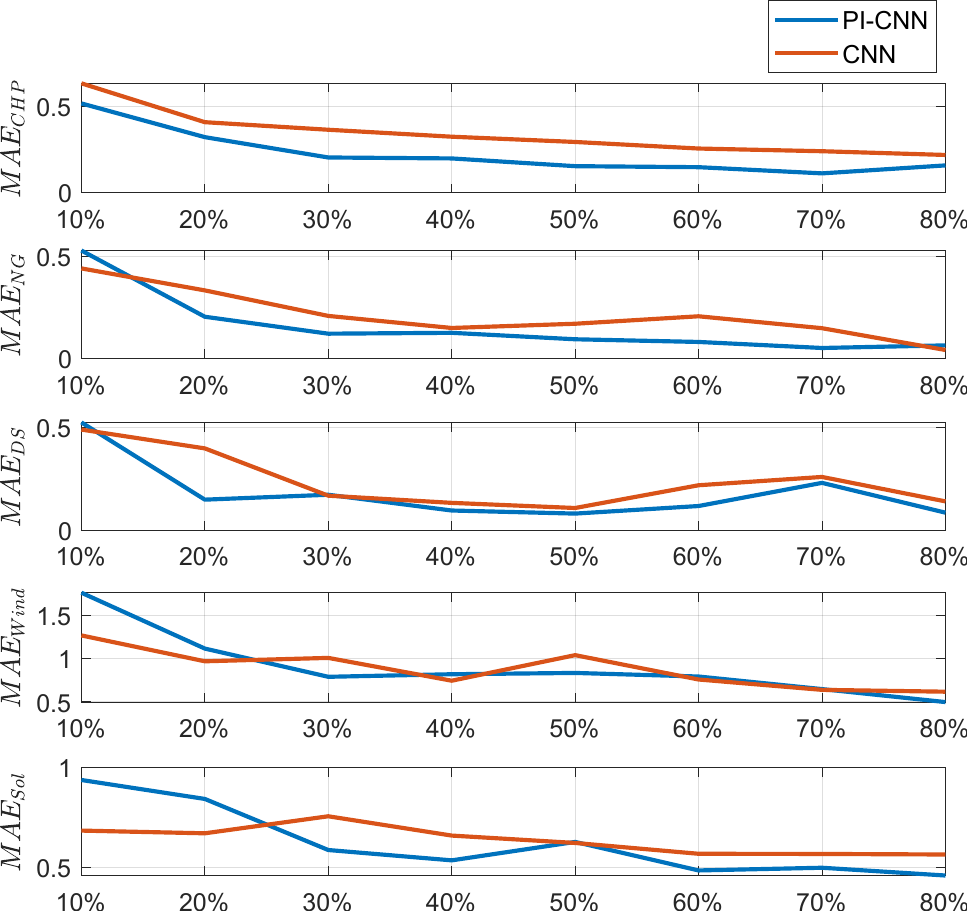}
}
\hspace{1cm} 
\subfloat[$R^2$ Comparison]{
    \includegraphics[scale=0.5]{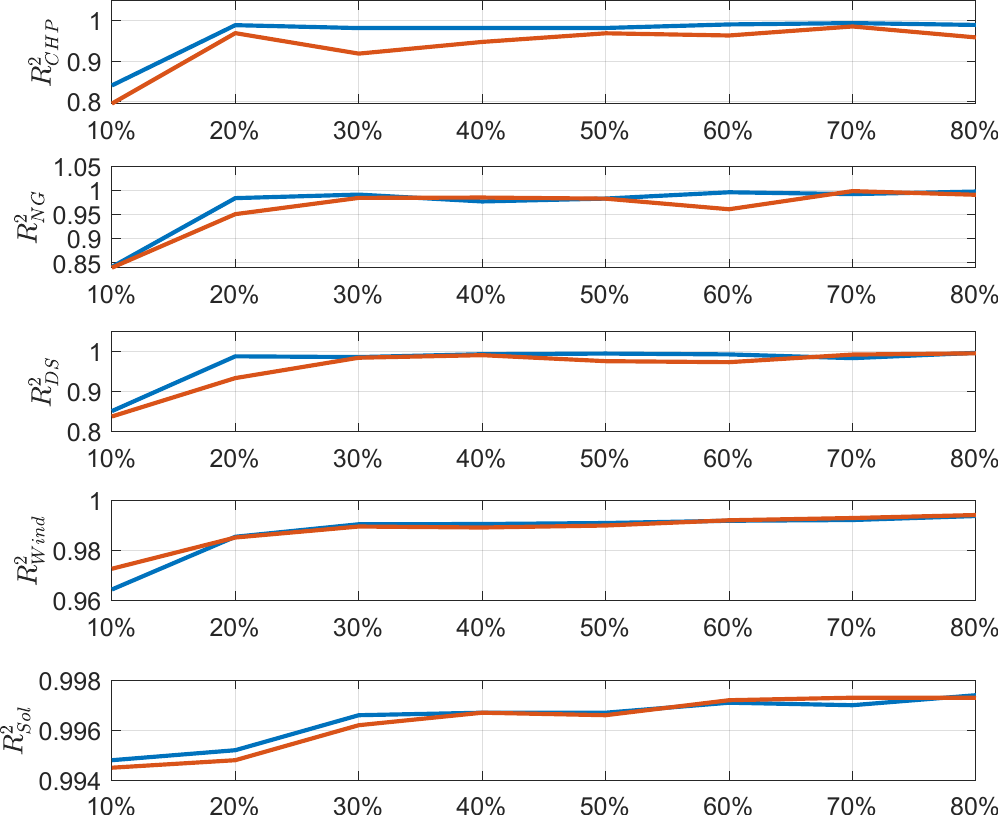}
}
\hspace{1cm} 
\subfloat[MSE Comparison]{
    \includegraphics[scale=0.5]{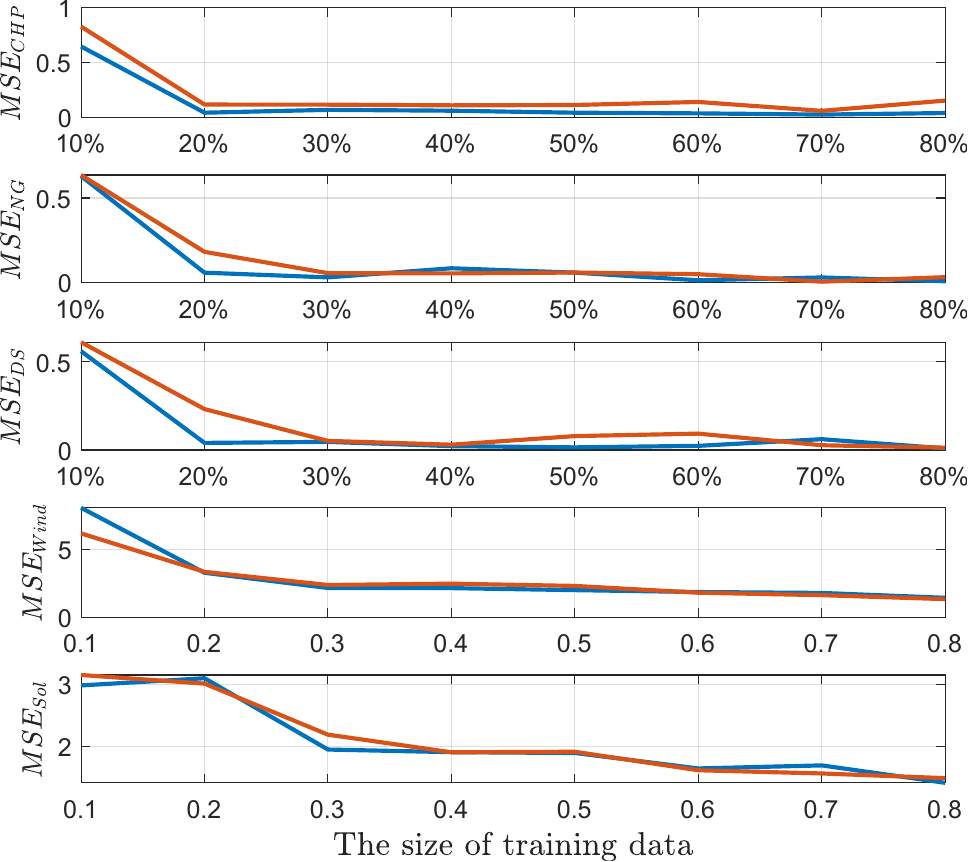}
}
\caption{Impact of size of training data on MSE, $R^2$, MAE values for CNN and PI-CNN}
\label{fig:MSE-MAE-value-comparison}
\end{figure}

\subsection{Training time and running time for all models}
This case study aims to provide a comprehensive comparison of the computational efficiency exhibited by the studied models. Therefore, we thoroughly investigated the training and running times of each model under consistent training conditions. Specifically, 80\% of the data was used for the training and 20\% of the data was used for testing in all models. In addition, all the algorithms were run for 150 training epochs. The results are presented in Table. \ref{tab:runtime_comparison}.

It is observed that when compared to the conventional CNN, the training time of the PI-CNN model is found to be 1.8 times longer under identical training conditions. This divergence in training times can be attributed to the increased computational complexity caused by the physical laws embedded in the loss function of PI-CNN, which can be regarded as a drawback of the PI-CNN model. However, in pursuit of superior prediction performance, sacrificing a small portion of training time can be deemed a better choice. Conversely, the training time of the DNN model remains nearly the same as the conventional CNN. In addition, although the training time of the PI-CNN is larger compared with the other two approaches, the runtimes of PI-CNN and CNN are fairly close (0.0414 and 0.0426 milliseconds, respectively) on the testing dataset, indicating no computational complexity for close to real-time execution of ED problem with PI-CNN. The DNN has the shorter runtime (0.0330 milliseconds), but given the lower accuracy of DNN compared with PI-CNN or CNN, one finds a trade-off between the runtime of the machine learning-based ED and accuracy of the prediction. Besides those, the prediction runtime for numerical optimization method(conventional ED) is 20 milliseconds. The runtime of numerical optimization is around 480 times slower than deep learning models, and the difference increases with the complexity/size of the system.

\begin{table}[]
\centering
\caption{Comparison of training time and runtime of
models}
\label{tab:runtime_comparison}
\begin{tabular}{ccccc}
\hline
Model             & PI-CNN  & CNN     & DNN      & Numerical\\ \hline
Training Times(s) & 640.297 & 349.495 & 345.402  & NA\\ \hline
Running Times(ms) & 0.0414  & 0.0426  & 0.0330   & 20\\ \hline
\end{tabular}
\end{table}

\begin{center}
\begin{table*}[]
\centering
\caption{Comparison between existing research on Learning-based ED in microgrid}
\label{tab:appcom}
\begin{tabular}{cccccc}
\hline
Reference & Approaches                                        & Deep Learning & Physics Inspired& Renewable Generation    & Best Results              \\ \hline
\cite{dong2021machine}     & Random Forest Regression        & No  & No  &  Wind,PV       & nRMSE:6.35\%                \\ \hline
\cite{kalakova2021novel}     & RBM   & Yes & No  & Wind    & MAPE:3.57\%                 \\ \hline
\cite{hafeez2020electric}         & FCRBM & Yes           & No                & None                    & MAPE:0.4525\%             \\ \hline
\cite{wen2019optimal}     & DRNN  & Yes & No  & PV      & MAE:4.369                   \\ \hline
\cite{gil2019economic}     & SPM & No  & No  &   None      & None Numerical Evaluation   \\ \hline
\cite{liu2022dynamic}        & Deep Reinforcement Learning                       & Yes           & No                & Wind,PV                 & None Numerical Evaluation \\ \hline
\cite{zhou2020combined}     & Deep Reinforcement Learning    & Yes & No  & Wind    &   None Numerical Evaluation  \\ \hline
\cite{ROCCHETTA2019291}         & ANN+Q-Learning             & Yes           & No                & 2 Renewable Generations & nRMSE:8.3\%               \\ \hline
\cite{du2019intelligent}     & Deep Neural Network            & Yes & No  & Wind    & Exponential Decay Rate:0.96 \\ \hline
Proposed Approach & PI-CNN                         & Yes & Yes & Wind, PV & MAE:0.0635  \\ \hline
\end{tabular}
\end{table*}
\end{center}

\subsection{Comparison with the State-of-the-Art Machine Learning-based ED}
This case study compares the existing research on machine learning-based microgrid ED with the proposed approach. Table.\ref{tab:appcom} illustrates the features of the existing approaches \cite{dong2021machine,kalakova2021novel,hafeez2020electric, wen2019optimal,gil2019economic,liu2022dynamic,zhou2020combined,ROCCHETTA2019291,du2019intelligent} for solving the ED problem in microgrid.

As it can be seen, existing approaches to address the uncertainties stemming from intermittent renewable generation in ED problems primarily rely on deep learning-based methods, while our proposed CNN approach is more accurate and requires less training epochs. In addition, none of these approaches utilize physics-inspired machine learning for solving the ED problem. In contrast, our proposed approach combines the physical laws with deep learning, presenting a novel contribution to the field. 

\section{Conclusion}
This paper presents a novel approach using physics-inspired convolutional neural network for solving the economic dispatch problem in Microgrids. By incorporating the constraints of a numerical economic dispatch problem into the learning process, the machine learning algorithm can be better trained to predict the solution of ED problem in real-time without solving the conventional computationally complex numerical optimization. The performance of the proposed approach is compared with DNN-based and conventional CNN-based approaches using several deterministic case studies. Results demonstrate that the PI-CNN approach  achieves excellent results with a small training dataset (even with 80\% less data) and runs very fast (0.04 milliseconds), which is extremely faster than conventional numerical optimization-based solutions (400 times faster). Future work will focus on enhancing the proposed design with a closed-loop economic dispatch engine that embeds dynamic models of assets to better respond to microgrid events.

\bibliographystyle{IEEEtran}
\bibliography{IEEEabrv,dp}

\begin{thebibliography}{10}
\providecommand{\url}[1]{#1}
\csname url@samestyle\endcsname
\providecommand{\newblock}{\relax}
\providecommand{\bibinfo}[2]{#2}
\providecommand{\BIBentrySTDinterwordspacing}{\spaceskip=0pt\relax}
\providecommand{\BIBentryALTinterwordstretchfactor}{4}
\providecommand{\BIBentryALTinterwordspacing}{\spaceskip=\fontdimen2\font plus
\BIBentryALTinterwordstretchfactor\fontdimen3\font minus \fontdimen4\font\relax}
\providecommand{\BIBforeignlanguage}[2]{{%
\expandafter\ifx\csname l@#1\endcsname\relax
\typeout{** WARNING: IEEEtran.bst: No hyphenation pattern has been}%
\typeout{** loaded for the language `#1'. Using the pattern for}%
\typeout{** the default language instead.}%
\else
\language=\csname l@#1\endcsname
\fi
#2}}
\providecommand{\BIBdecl}{\relax}
\BIBdecl

\bibitem{wang2022multi}
L.~Wang, X.~An, H.~Xu, and Y.~Zhang, ``Multi-agent-based collaborative regulation optimization for microgrid economic dispatch under a time-based price mechanism,'' \emph{Electric Power Systems Research}, vol. 213, p. 108760, 2022.

\bibitem{ed1}
M.~A. Velasquez, J.~Barreiro-Gomez, N.~Quijano, A.~I. Cadena, and M.~Shahidehpour, ``Intra-hour microgrid economic dispatch based on model predictive control,'' \emph{IEEE Transactions on Smart Grid}, vol.~11, no.~3, pp. 1968--1979, 2019.

\bibitem{obama2017irreversible}
B.~Obama, ``The irreversible momentum of clean energy,'' \emph{Science}, vol. 355, no. 6321, pp. 126--129, 2017.

\bibitem{wu2022data}
J.~Wu, Y.~Liu, X.~Chen, C.~Wang, and W.~Li, ``Data-driven adjustable robust day-ahead economic dispatch strategy considering uncertainties of wind power generation and electric vehicles,'' \emph{International Journal of Electrical Power \& Energy Systems}, vol. 138, p. 107898, 2022.

\bibitem{two-stage}
\BIBentryALTinterwordspacing
Y.~Li, J.~Wang, D.~Zhao, G.~Li, and C.~Chen, ``A two-stage approach for combined heat and power economic emission dispatch: Combining multi-objective optimization with integrated decision making,'' \emph{Energy (Oxford)}, vol. 162, 11 2018. [Online]. Available: \url{https://www.osti.gov/biblio/1490180}
\BIBentrySTDinterwordspacing

\bibitem{ed2}
H.~Hou, M.~Xue, Y.~Xu, Z.~Xiao, X.~Deng, T.~Xu, P.~Liu, and R.~Cui, ``Multi-objective economic dispatch of a microgrid considering electric vehicle and transferable load,'' \emph{Applied Energy}, vol. 262, p. 114489, 2020.

\bibitem{hybrid}
\BIBentryALTinterwordspacing
H.~Narimani, S.-E. Razavi, A.~Azizivahed, E.~Naderi, M.~Fathi, M.~H. Ataei, and M.~R. Narimani, ``A multi-objective framework for multi-area economic emission dispatch,'' \emph{Energy}, vol. 154, pp. 126--142, 2018. [Online]. Available: \url{https://www.sciencedirect.com/science/article/pii/S036054421830687X}
\BIBentrySTDinterwordspacing

\bibitem{ed3}
Z.~Zhang, D.~Yue, C.~Dou, and H.~Zhang, ``Multiagent system-based integrated design of security control and economic dispatch for interconnected microgrid systems,'' \emph{IEEE Transactions on Systems, Man, and Cybernetics: Systems}, vol.~51, no.~4, pp. 2101--2112, 2020.

\bibitem{Stochastic}
Y.-Y. Lee and R.~Baldick, ``A frequency-constrained stochastic economic dispatch model,'' \emph{IEEE Transactions on Power Systems}, vol.~28, no.~3, pp. 2301--2312, 2013.

\bibitem{ed4}
Y.~Xu, Z.~Dong, Z.~Li, Y.~Liu, and Z.~Ding, ``Distributed optimization for integrated frequency regulation and economic dispatch in microgrids,'' \emph{IEEE Transactions on Smart Grid}, vol.~12, no.~6, pp. 4595--4606, 2021.

\bibitem{mean-tracking}
\BIBentryALTinterwordspacing
Z.~Lin, H.~Chen, Q.~Wu, W.~Li, M.~Li, and T.~Ji, ``Mean-tracking model based stochastic economic dispatch for power systems with high penetration of wind power,'' \emph{Energy}, vol. 193, p. 116826, 2020. [Online]. Available: \url{https://www.sciencedirect.com/science/article/pii/S0360544219325216}
\BIBentrySTDinterwordspacing

\bibitem{ed5}
B.~Huang, L.~Liu, H.~Zhang, Y.~Li, and Q.~Sun, ``Distributed optimal economic dispatch for microgrids considering communication delays,'' \emph{IEEE Transactions on Systems, Man, and Cybernetics: Systems}, vol.~49, no.~8, pp. 1634--1642, 2019.

\bibitem{shuai2018stochastic}
H.~Shuai, J.~Fang, X.~Ai, Y.~Tang, J.~Wen, and H.~He, ``Stochastic optimization of economic dispatch for microgrid based on approximate dynamic programming,'' \emph{IEEE Transactions on Smart Grid}, vol.~10, no.~3, pp. 2440--2452, 2018.

\bibitem{yeh2020new}
W.-C. Yeh, M.-F. He, C.-L. Huang, S.-Y. Tan, X.~Zhang, Y.~Huang, and L.~Li, ``New genetic algorithm for economic dispatch of stand-alone three-modular microgrid in dongao island,'' \emph{Applied Energy}, vol. 263, p. 114508, 2020.

\bibitem{hou2020multi}
H.~Hou, M.~Xue, Y.~Xu, Z.~Xiao, X.~Deng, T.~Xu, P.~Liu, and R.~Cui, ``Multi-objective economic dispatch of a microgrid considering electric vehicle and transferable load,'' \emph{Applied Energy}, vol. 262, p. 114489, 2020.

\bibitem{dong2021machine}
W.~Dong, Q.~Yang, W.~Li, and A.~Y. Zomaya, ``Machine-learning-based real-time economic dispatch in islanding microgrids in a cloud-edge computing environment,'' \emph{IEEE Internet of Things Journal}, vol.~8, no.~17, pp. 13\,703--13\,711, 2021.

\bibitem{kalakova2021novel}
A.~Kalakova, H.~K. Nunna, P.~K. Jamwal, and S.~Doolla, ``A novel genetic algorithm based dynamic economic dispatch with short-term load forecasting,'' \emph{IEEE Transactions on Industry Applications}, vol.~57, no.~3, pp. 2972--2982, 2021.

\bibitem{hafeez2020electric}
G.~Hafeez, K.~S. Alimgeer, and I.~Khan, ``Electric load forecasting based on deep learning and optimized by heuristic algorithm in smart grid,'' \emph{Applied Energy}, vol. 269, p. 114915, 2020.

\bibitem{wen2019optimal}
L.~Wen, K.~Zhou, S.~Yang, and X.~Lu, ``Optimal load dispatch of community microgrid with deep learning based solar power and load forecasting,'' \emph{Energy}, vol. 171, pp. 1053--1065, 2019.

\bibitem{gil2019economic}
W.~Gil-Gonz{\'a}lez, O.~D. Montoya, E.~Holgu{\'\i}n, A.~Garces, and L.~F. Grisales-Nore{\~n}a, ``Economic dispatch of energy storage systems in dc microgrids employing a semidefinite programming model,'' \emph{Journal of Energy Storage}, vol.~21, pp. 1--8, 2019.

\bibitem{liu2022dynamic}
Z.~Liu, Y.~Liu, H.~Xu, S.~Liao, K.~Zhu, and X.~Jiang, ``Dynamic economic dispatch of power system based on ddpg algorithm,'' \emph{Energy Reports}, vol.~8, pp. 1122--1129, 2022.

\bibitem{zhou2020combined}
S.~Zhou, Z.~Hu, W.~Gu, M.~Jiang, M.~Chen, Q.~Hong, and C.~Booth, ``Combined heat and power system intelligent economic dispatch: A deep reinforcement learning approach,'' \emph{International journal of electrical power \& energy systems}, vol. 120, p. 106016, 2020.

\bibitem{ROCCHETTA2019291}
\BIBentryALTinterwordspacing
R.~Rocchetta, L.~Bellani, M.~Compare, E.~Zio, and E.~Patelli, ``A reinforcement learning framework for optimal operation and maintenance of power grids,'' \emph{Applied Energy}, vol. 241, pp. 291--301, 2019. [Online]. Available: \url{https://www.sciencedirect.com/science/article/pii/S0306261919304222}
\BIBentrySTDinterwordspacing

\bibitem{du2019intelligent}
Y.~Du and F.~Li, ``Intelligent multi-microgrid energy management based on deep neural network and model-free reinforcement learning,'' \emph{IEEE Transactions on Smart Grid}, vol.~11, no.~2, pp. 1066--1076, 2019.

\bibitem{jia2022convopf}
Y.~Jia, X.~Bai, L.~Zheng, Z.~Weng, and Y.~Li, ``Convopf-dop: A data-driven method for solving ac-opf based on cnn considering different operation patterns,'' \emph{IEEE Transactions on Power Systems}, vol.~38, no.~1, pp. 853--860, 2022.

\bibitem{karniadakis2021physics}
G.~E. Karniadakis, I.~G. Kevrekidis, L.~Lu, P.~Perdikaris, S.~Wang, and L.~Yang, ``Physics-informed machine learning,'' \emph{Nature Reviews Physics}, vol.~3, no.~6, pp. 422--440, 2021.

\bibitem{harbola2019one}
S.~Harbola and V.~Coors, ``One dimensional convolutional neural network architectures for wind prediction,'' \emph{Energy Conversion and Management}, vol. 195, pp. 70--75, 2019.

\bibitem{moazeni2020maximizing}
F.~Moazeni, J.~Khazaei, and J.~P.~P. Mendes, ``Maximizing energy efficiency of islanded micro water-energy nexus using co-optimization of water demand and energy consumption,'' \emph{Applied Energy}, vol. 266, p. 114863, 2020.

\bibitem{hijjo2017pv}
M.~Hijjo, F.~Felgner, and G.~Frey, ``Pv-battery-diesel microgrid layout design based on stochastic optimization,'' in \emph{2017 6th International Conference on Clean Electrical Power (ICCEP)}.\hskip 1em plus 0.5em minus 0.4em\relax IEEE, 2017, pp. 30--35.

\bibitem{moazeni2021step}
F.~Moazeni, J.~Khazaei, and A.~Asrari, ``Step towards energy-water smart microgrids; buildings thermal energy and water demand management embedded in economic dispatch,'' \emph{IEEE Transactions on Smart Grid}, vol.~12, no.~5, pp. 3680--3691, 2021.

\bibitem{krizhevsky2017imagenet}
A.~Krizhevsky, I.~Sutskever, and G.~E. Hinton, ``Imagenet classification with deep convolutional neural networks,'' \emph{Communications of the ACM}, vol.~60, no.~6, pp. 84--90, 2017.

\bibitem{afrasiabi2019probabilistic}
M.~Afrasiabi, M.~Mohammadi, M.~Rastegar, and A.~Kargarian, ``Probabilistic deep neural network price forecasting based on residential load and wind speed predictions,'' \emph{IET Renewable Power Generation}, vol.~13, no.~11, pp. 1840--1848, 2019.

\bibitem{kiranyaz20211d}
S.~Kiranyaz, O.~Avci, O.~Abdeljaber, T.~Ince, M.~Gabbouj, and D.~J. Inman, ``1d convolutional neural networks and applications: A survey,'' \emph{Mechanical systems and signal processing}, vol. 151, p. 107398, 2021.

\bibitem{currie2012opti}
J.~Currie, D.~I. Wilson, N.~Sahinidis, and J.~Pinto, ``Opti: Lowering the barrier between open source optimizers and the industrial matlab user,'' \emph{Foundations of computer-aided process operations}, vol.~24, p.~32, 2012.

\end{thebibliography}

\end{document}